\documentclass{osa-article}

\journal{oe}
\articletype{Research Article}
\begin{document}

\title{Efficient computation of backprojection arrays for 3D light field deconvolution}

\author{Martin Eberhart\authormark{1}}

\address{\authormark{1}High Enthalpy Flow Diagnostics Group, Institute of Space Systems, University of Stuttgart, Pfaffenwaldring 29, 70569 Stuttgart, Germany}

\email{\authormark{*}eberhart@irs.uni-stuttgart.de} %% email address is required



\begin{abstract}
Light field deconvolution allows three-dimensional investigations from a single snapshot recording of a plenoptic camera. It is based on a linear image formation model, and iterative volume reconstruction requires to define the backprojection of individual image pixels into object space. This is effectively a reversal of the point spread function (PSF), and backprojection arrays $\mathbf{H'}$ can be derived from the shift-variant PSFs $\mathbf{H}$ of the optical system, which is a very time consuming step for high resolution cameras. This paper illustrates the common structure of backprojection arrays and the significance of their efficient computation. A new algorithm is presented to determine $\mathbf{H'}$ from $\mathbf{H}$, which is based on the distinct relation of the elements' positions within the two multi-dimensional arrays. It permits a pure array re-arrangement, and while results are identical to those from published codes, computation times are drastically reduced. This is shown by benchmarking the new method using various sample PSF arrays against existing algorithms. The paper is complemented by practical hints for the experimental acquisition of light field PSFs in a photographic setup.
\end{abstract}

%%%%%%%%%%%%%%%%%%%%%%%%%%  body  %%%%%%%%%%%%%%%%%%%%%%%%%%
\section{Introduction}
\begin{figure}[b]
	\centering
	\includegraphics[width=\linewidth]{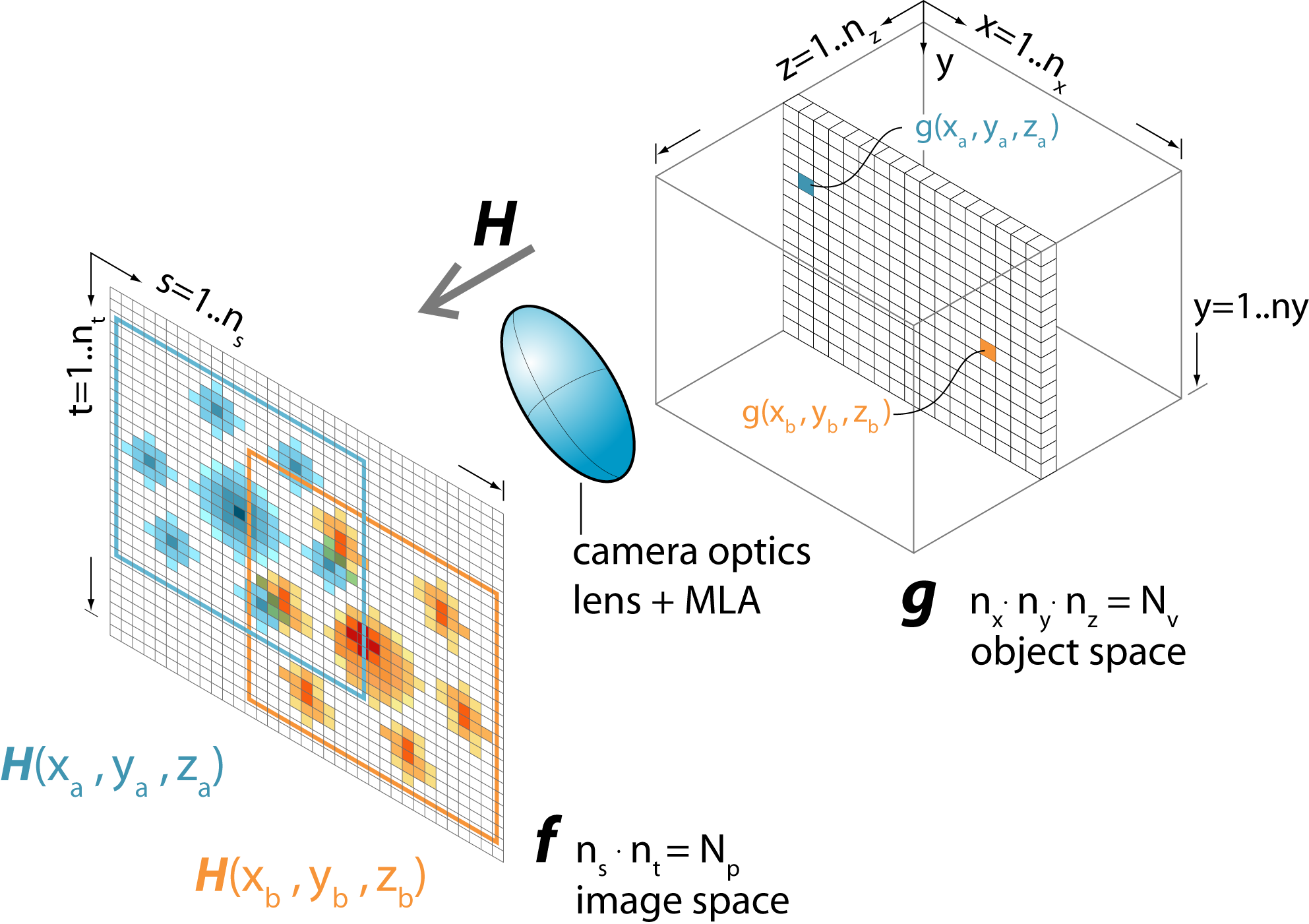}
	\caption{Sketch of image space, object space and definition of the used dimensions and indices.}
	\label{fig:sketch}
\end{figure}

Following the seminal publication by \textsc{Levoy} et al.~\cite{Levoy_2006_01}, a considerable amount of work has been dedicated to the development of 3D diagnostics for microscopes, using digital plenoptic cameras as imaging devices. Compared to standard photographic cameras, these instruments allow to measure not only the intensity distribution at their image plane, but also record additional directional information on the light rays in the scene. This is realized by inserting an array of microlenses (MLA) into the optical path, close to the image sensor, which distributes light according to its direction onto different pixels. The captured image is therefore a coded representation of the 4D light field, and its spatio-angular information allows to derive depth coordinates, and, for transparent objects, the 3D intensity distribution within the volume. This is a scanless technique based on a single snapshot recording, with the attractive potential of investigating dynamic processes in three dimensions.\\
In its original implementation, light field microscopy mimicked the work flow of traditional 3D deconvolution: Here the input data is in the form of a focal stack, recorded as a sequence while sweeping the object along the optical axis~\cite{McNally_1999_01}. 
Image formation is modeled as a convolution of the object space intensity distribution with the point spread function (PSF) of the microscope, which defines light transport within the optical system. Deconvolution methods seek to revert this process,
using a known PSF as a tool for estimating the original volume from recorded image stacks. 
With captured light field data, however, focal stacks can be computed from a single exposure by synthetic refocusing~\cite{Isaksen_2000_01}, eliminating the time-consuming acquisition of sequences. Capturing of transient processes is therefore only limited by the frame rate of the camera. This comes at a cost:
As a plenoptic camera typically has to sacrifice lateral resolution for the additional directional information, early light field microscopy suffered from comparably low pixel counts of the reconstructed volumes.\\
A significant improvement was published by Broxton et al.~\cite{Broxton_2013_01}, which circumvented the generation of focal stacks, but instead directly operated on light field data recorded by a plenoptic camera. So-called light field deconvolution is closely related to super-resolution approaches~\cite{Bishop_2012_01}, and models the formation of the camera's raw images, i.e. the captured 4D light field, by applying a measurement matrix $\mathbf{H}$ to the intensity distribution within object space. For a discretized volume, this matrix, also termed \textit{light field PSF}, defines the intensity received by each single pixel from all individual light-emitting voxels. These interrelations are sketched in Fig.~\ref{fig:sketch}, which also serves as a reference for the definitions and notations that are going to be used throughout the paper. Similar
to a standard PSF, $\mathbf{H}$ is found by measuring or simulating the sensor's pixel response to a subresolution, point-like light source at all voxel positions. The sketch shows light emitting voxels at two distinct positions within the object space $\mathbf{g}$, and the resulting pixel intensities within the image space $\mathbf{f}$.\\
This technique provided a considerable boost in volume resolution and allowed the 3D analysis of live animals under a microscope~\cite{Prevedel_2014_01}. However, resolution was non-uniform across the depth of field and suffered from artifacts especially close to the native object plane. This could be ameliorated by using an enhanced optical setup with additional phase plates, as suggested by \textsc{Cohen} et al.~\cite{Cohen_2014_01}. A different route was taken by \textsc{Stefanoiu} et al.~\cite{Stefanoiu_2019_01}, who incorporated a depth-dependent filtering in the light field deconvolution algorithm, which resulted in an artifact-free volume reconstruction. A significant computational speed-up and reduction of artifacts was achieved by \textsc{Lu} et al.~\cite{Lu_2019_01} by transferring both the PSF $\mathbf{H}$ and the measured light field data into phase space and separating the spatial frequencies in the deconvolution step.\\
At the core of these techniques are deconvolution algorithms, which have been in use before, performing de-blurring of two-dimensional images as well as 3D reconstruction from traditional microscopic focal stacks. All of the above mentioned publications rely on the classical Richardson-Lucy scheme, but a variety of other algorithms may be suitable as well, with an overview given e.g. by \textsc{Sage} et al.~\cite{Sage_2017_01}. In addition to the PSF $\mathbf{H}$, most of these algorithms require to define light transport through the optical system in the reversed direction: While $\mathbf{H}$ describes a forward projection of the volume onto the image plane, we need to formulate a backprojection $\mathbf{H'}$ of the pixels into object space. 
It is, however, not trivial to derive this backprojection from a given light field PSF, because the latter is shift-variant, which means that point sources at different voxel positions create different sensor responses, which are commonly stored in a multi-dimensional array $\mathbf{H}$.
While the literature gives detailed explanations of the wave optical modeling of the light field PSF, it does not provide recipes for the computation of $\mathbf{H'}$ from $\mathbf{H}$, which requires an elaborate procedure. The publications~\cite{Prevedel_2014_01, Stefanoiu_2019_01, Lu_2019_01} include computer code that also performs this step, but without specific comments on its functionality. Moreover, computation of $\mathbf{H'}$ using these codes is slow for modern plenoptic cameras with a high pixel count and MLAs in a hexagonal grid.\\
The contribution of this paper is the following: It first clarifies the significance and structure of $\mathbf{H'}$ and its relation to the original light field PSF, and complements this by hints on how to record the PSF experimentally in a suitable array. The paper illustrates the relevance of a quick computation of $\mathbf{H'}$, and then presents a simple and rapid algorithm for this task. It is purely based on the distinct relation of the elements' positions within $\mathbf{H}$ and $\mathbf{H'}$, respectively, and a benchmark using sample PSFs shows a substantial speed-up compared to previously published codes while providing identical results.

\section{Image formation and recording of the light field}
\begin{figure}[b]
	\centering
	\includegraphics[width=0.6\linewidth]{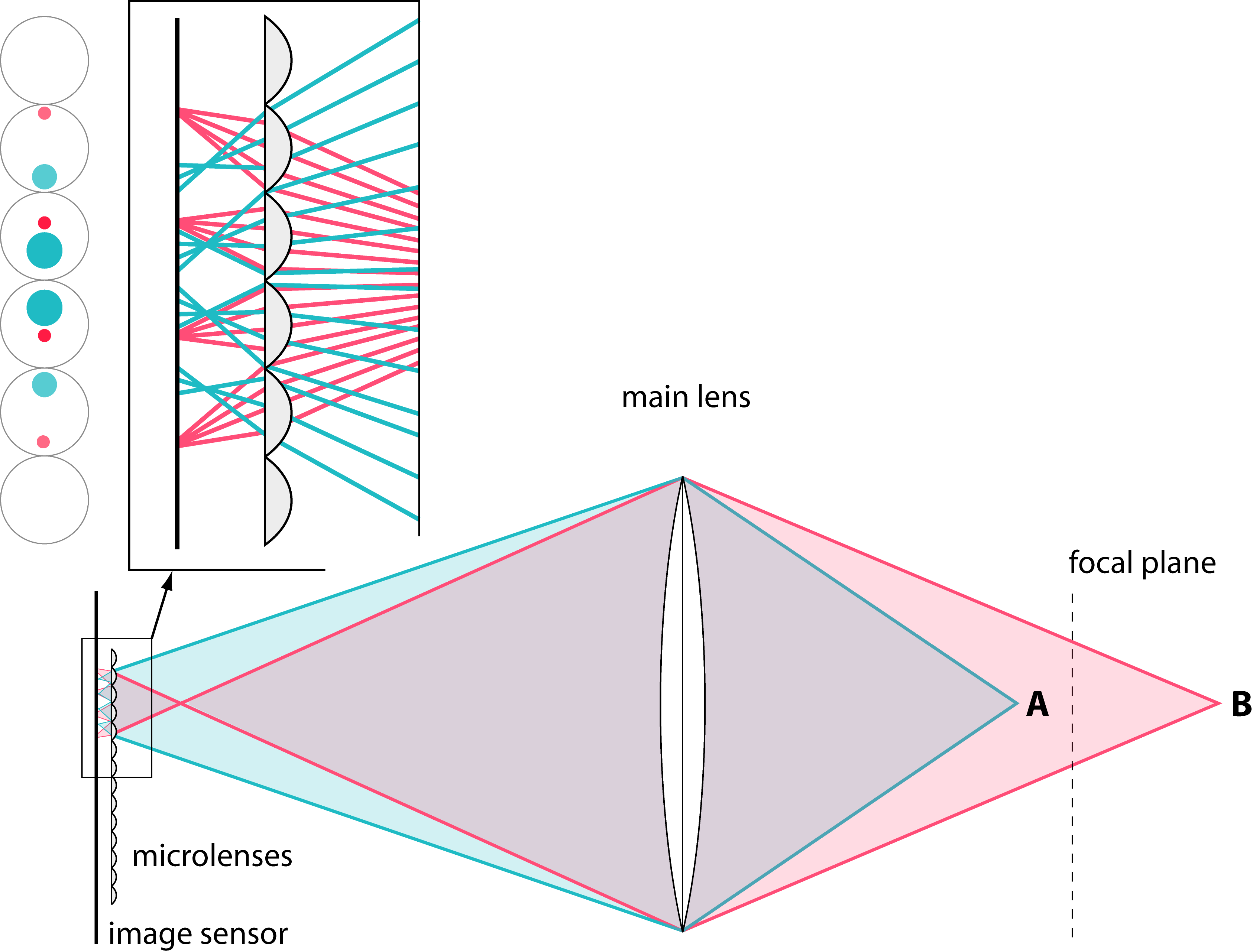}
	\caption{Sketch of a general plenoptic camera and formation of pixel patterns at the image plane by recording single points in object space.}
	\label{fig:pcam}
\end{figure}

\begin{figure}[t]
	\centering
	\includegraphics[width=\linewidth]{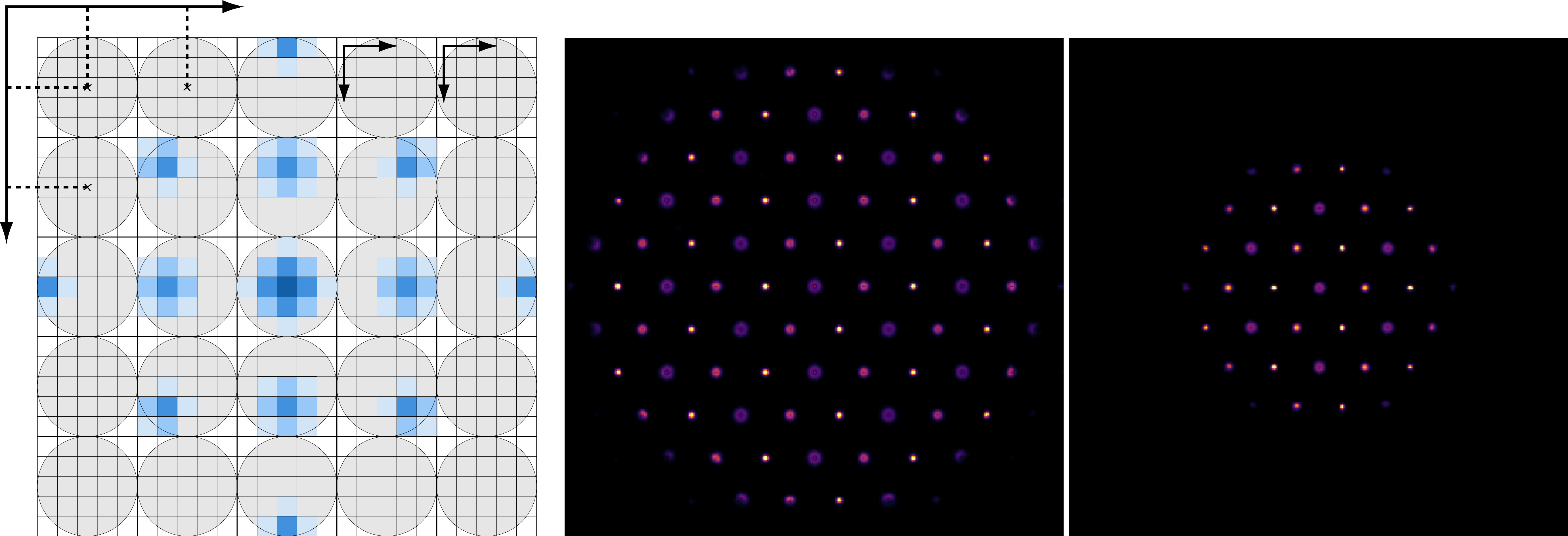}
	\caption{\textit{Left:} Schematic of a plenoptic raw image, composed of multiple circular microimages (gray circles). It can be interpreted as an overlay of two two-dimensional coordinate systems and is a recording of the 4D light field. \textit{Middle/Right:} Intensity distributions produced by a single point in space at different depth positions,
	recorded by a plenoptic camera. These patterns define point spread functions (PSF). The microlens array consists of three different lens types, which is obvious in the image.}
	\label{fig:MLA_KOS}
\end{figure}

In a standard photographic camera, single sensor pixels (or grains of a chemical film) integrate the incident light over a certain solid angle. As a consequence, directional information is lost and the images are flat. A plenoptic camera uses a microlens array (MLA) close to the sensor, as sketched in Fig.~\ref{fig:pcam}, which distributes light rays according to their direction onto different sensor pixels~\cite{Adelson_1992_01}. The MLA has the effect of a multiplexer~\cite{Wetzstein_2013_01}, coding the lost directional information into the captured raw image, which is a recording of the 4D light field of the scene. Two points in object space, marked A and B in the figure, generate distinct spot patterns on the image sensor. \\
These patterns are representations of the light field, often termed plenoptic function, which defines the transport of light energy along rays in space. In its simplest form this is a 4D function, with two spatial coordinates describing position, and two angular coordinates defining orientation. Consequently, the raw image recorded by a plenoptic camera can be interpreted as an overlay of two two-dimensional coordinate systems, sketched on the left of Fig.~\ref{fig:MLA_KOS}. 
The gray circles illustrate circular microimages formed behind the lenslets of the MLA, and their position is expressed by an outer coordinate system. Little boxes represent sensor pixels, and their position within the respective microimages is given by an inner system. The exact physical meaning of these systems depends on the type of plenoptic camera: In its initial implementation, sampling of spatial and angular data is strictly separated, with the outer and inner coordinate system defining position and ray orientation, respectively~\cite{Ng_2006_01}.
Angular resolution of the sampled light field is defined by the number of pixels within a microimage, and lateral resolution of the images is given by the number of microlenses. This illustrates the already mentioned trade-off of a plenoptic camera, sacrificing pixel resolution for additional directional information. For a so-called focused plenoptic camera, with a different separation between MLA and sensor, sampling of spatial and directional data is intertwined~\cite{Lumbsdaine_2009_01}. This allows an increased lateral resolution at the cost of a more complex processing of the raw data.
Regardless of the type of camera, a sensor recording of a single light point, as outlined in Fig.~\ref{fig:pcam}, represents a point spread function, a spatio-angular light field PSF, that defines the optical system. Such a PSF may look like in the sketch on the left of Fig.~\ref{fig:MLA_KOS}. The right hand side of the figure shows actual PSF patterns captured experimentally with a plenoptic camera for points at two different depth positions. Here a so-called multi-focus camera was used, which features an MLA with three different lens types in a hexagonal arrangement~\cite{Georgiev_2012_01,Perwass_2012_01}. \\
The pixel patterns formed by discrete light points are, however, not only a function of the point's depth coordinate, but also depend on its lateral position. This holds true even if the plenoptic camera is used at a microscope with a telecentric objective lens. A single PSF is therefore not sufficient to characterize light transport within the system, but a suitable set of PSF patterns must be used for light field deconvolution. \\

\subsection{PSF structure}

\begin{figure}[h]
	\centering
	\includegraphics[width=0.9\linewidth]{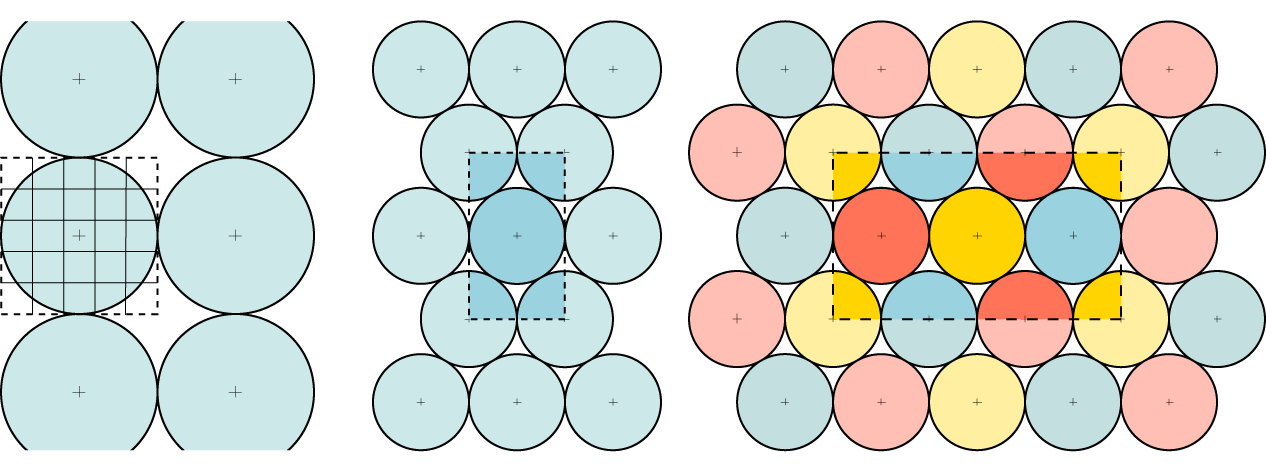}
	\caption{Sketch of different MLA layouts. \textit{Left:} Rectangular grid, \textit{Middle:} Hexagonal grid, \textit{Right:} Hexagonal grid and three different lenslet types, mounted e.g. in a Raytrix R29 camera. Dashed boxes mark representative regions or elementary cells.}
	\label{fig:MLA-box}
\end{figure}

The coordinate system used in this paper and various dimensions and indices are included in the sketch of
Fig.~\ref{fig:sketch}. The image $\mathbf{f}$ is formed on an $s,t$-plane and is discretized into $n_s \times n_t$ pixels. Note that here the superposition of two coordinate systems is disregarded, but instead absolute pixel positions are used. The volume $\mathbf{g}$ is contained within an $x,y,z$-cube, with the z-axis aligned to the camera's optical axis. It is discretized accordingly into $n_x \times n_y \times n_z$ voxels. Lateral voxel sizes are tied to the physical pixel pitch of the sensor by the main lens magnification, while voxel depth is arbitrary and subject to the chosen discretization. Light emitted by a single point (or voxel) in space is transferred through object space and the optical system and is then recorded by the sensor pixels as an intensity distribution, as shown in Fig.~\ref{fig:sketch} in blue and red.
A complete light field PSF $\mathbf{H}$ is therefore a collection of these patterns, captured with a light point at all relevant (voxel-) positions within object space. In its full extend, $\mathbf{H}$ establishes a relation between light emitted at each voxel and the received intensity at each pixel. This is a huge amount of data which can hardly be handled by a processing algorithm. Two findings help to drastically cut required information~\cite{Broxton_2013_01}: Point spread patterns of single points are very sparse, so that zero pixels can be largely discarded. This means cropping recorded patterns using rectangular cutouts, illustrated by colored boxes in Fig.~\ref{fig:sketch}. And second, the regular arrangement of the lenslets within the MLA results in periodically repeating pixel patterns when shifting a light point in lateral directions. For each axial depth, the optical system is therefore defined by a limited number of light field patterns, which need to be determined with a point source at all voxel positions within a representative area, called an elementary cell. Shape and size of this area depend on the MLA layout, as illustrated in Fig.~\ref{fig:MLA-box}.
For lenslets arranged in a rectangular grid, on the left of the figure, the elementary cell is simply quadratic, covering the extends of a microimage. The light field PSF $\mathbf{H}$ has to hold sensor responses to a point source placed at all corresponding voxel positions, which is 5$\times$5 in this example, the number of pixels within a microimage. Absolute physical voxel sizes in object space depend on the magnification of the main lens, subject to optional super- or undersampling factors in the code implementation. For a hexagonal MLA layout, shown in the middle of Fig.~\ref{fig:MLA-box}, the elementary cell takes the shape of the dashed box, and shifting the point source beyond the corresponding limits in object space will result in repeating patterns. For the case of a multi-focus camera with three different microlenses, sketched on the right, the elementary cell has to be expanded. It is the smallest possible area of the MLA layout that allows to tile the entire plane with periodic copies. Considering lines of symmetry can help to further reduce its size.\\
This approach is based on the assumption of a shift-invariant PSF of the isolated main lens, which holds, strictly speaking, only for ideal telecentric systems like microscopes~\cite{Levoy_2006_01}. If a plenoptic camera is used in a photographic setup, this assumption may be at stake and must be carefully examined. The general concept of deriving a backprojection array from a given light field PSF is, however, not limited to orthographic projections in microscope systems.\\
With this background it is clear that image formation on the sensor plane of a plenoptic camera cannot be modeled as a convolution, because the forward projection, defined by $\mathbf{H}$, varies from point to point. In a discretized form, it is given by the generalized linear equation~\cite{Broxton_2013_01}
\begin{equation}
	\mathbf{f} = \mathbf{H} \, \, \mathbf{g}
	\label{eq:iformation}
\end{equation}
Image formation can be interpreted as distributing copies of the individual light field patterns of $\mathbf{H}$, scaled by the magnitude of the respective object space voxels in $\mathbf{g}$, which sum up to form the light field image $\mathbf{f}$. In computer code, this distribution can be efficiently implemented as single, discrete convolutions. To be able to do so, it has to be ensured that the patterns are mapped to the correct pixel positions: During acquisition of $\mathbf{H}$, the bounds of the cutouts, drawn as colored boxes in Fig.~\ref{fig:sketch}, must be shifted likewise when the point source traverses from voxel to voxel.\\
The light field PSF $\mathbf{H}$ has to be given in a structure that is suitable for actual computer applications. Codes published in~\cite{Prevedel_2014_01,Stefanoiu_2019_01,Lu_2019_01} conveniently define $\mathbf{H}$ as a 5-dimensional array, with two dimensions (s,t) holding the image pixels, and three dimensions (x,y,z) defining the respective object space coordinates. It is therefore a collection of two-dimensional patterns, arranged in a three-dimensional array.\\

\subsection{PSF acquisition}
\begin{figure}[b]
	\centering
	\includegraphics[width=0.5\linewidth]{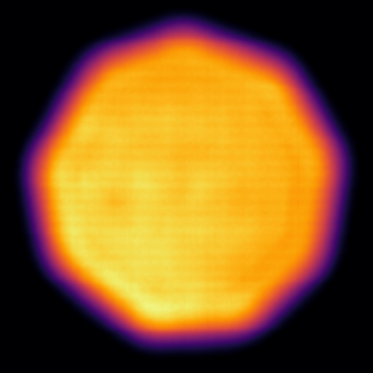}
	\caption{Summation of all elements of the PSF matrix $\mathbf{H}$ within one depth plane. The PSF was recorded experimentally using a photographic main lens, and the structure of the aperture constructed from 9 movable blades is clearly seen.   }
	\label{fig:sumpsf}
\end{figure}
For microscope applications, determination of the light field PSF $\mathbf{H}$ has commonly been done on a theoretical basis~\cite{Broxton_2013_01,Prevedel_2014_01,Cohen_2014_01,Stefanoiu_2019_01}. This is reasonable, because the optical system is very precise, and due to the high magnification, the required shifts of the point source are on an extremely small scale. For photographic setups, however, it is beneficial to cast this into an experimental calibration procedure to account for imperfections and not precisely defined elements. This requires a very small, point-like light source, close to an isotropic emitter, that is precisely positioned within the calibration volume by a computerized 3-axes translation stage~\cite{Eberhart_2021_01}. The stage is synchronized with the plenoptic camera, which records light field raw data at each step, and the cutout pixel patterns are then forged into the array $\mathbf{H}$. Here the cone-like field of view of the photographic main lens needs to be considered, with the consequence that the physical boundaries of the elementary cells vary with the depth coordinate.\\It
is interesting to examine the summation of all the elements of $\mathbf{H}$ within one depth plane, shown in Fig.~\ref{fig:sumpsf}. Here the PSF was acquired experimentally with a photographic plenoptic camera (Raytrix R29 with Nikkor 200mm f/4 main lens)~\cite{Eberhart_2021_02}. By subsequently shifting point source and cutout region, the aperture of the main lens is gradually illuminated from all possible directions, filling up the spatio-angular data in the image. This reveals the complex shape of the aperture, designed with 9 movable blades, which are variable and not precisely know, showcasing the requirement of an experimental calibration. \\

\section{Deconvolution and backprojection array}

Volume reconstruction is an inverse problem that seeks to revert Eq.~(\ref{eq:iformation}), trying to find the original volume $\mathbf{g}$ from a measured light field $\mathbf{f}$, using a known PSF  $\mathbf{H}$. This is an ill-posed task, and inevitable noise prevents a simple inversion of the image formation equation, so that iterative deconvolution techniques have to be applied instead. One representative of the wide range of different algorithms is the classical Richardson-Lucy scheme, which is still common in advanced deconvolution approaches~\cite{Shajkofci_2020_01}. It has so far been used in the framework of light field deconvolution, and its iterative update in matrix-vector notation reads~\cite{Stefanoiu_2019_01}
\begin{equation}
	\mathbf{g}^{(k+1)} =  \frac{\mathbf{g}^{(k)}}{   \mathbf{H'} \, \mathbf{1}} \left[  \mathbf{H'} \frac{ \, \mathbf{f}}{ \mathbf{H} \, \mathbf{g}^{(k)}} \right]
	\label{eq:RL}
\end{equation}
Note that this must not be interpreted as standard matrix multiplications, as the PSF is in the form of a multi-dimensional array, which has to be implemented accordingly in the computer code. In essence, the scheme of Eq.~(\ref{eq:RL}) computes an error quotient by comparing the measured image $\mathbf{f}$ to the forward projection of the current volume estimate, $\mathbf{H} \, \mathbf{g}^{(k)}$. This error is then backprojected into object space by means of the array $\mathbf{H'}$, and updates the volumetric intensity distribution $\mathbf{g}$.\\
This paper now seeks to establish a relation between the light field PSF $\mathbf{H}$ and the backprojection array $\mathbf{H'}$.
\begin{figure}[b]
	\centering
	\includegraphics[width=\textwidth]{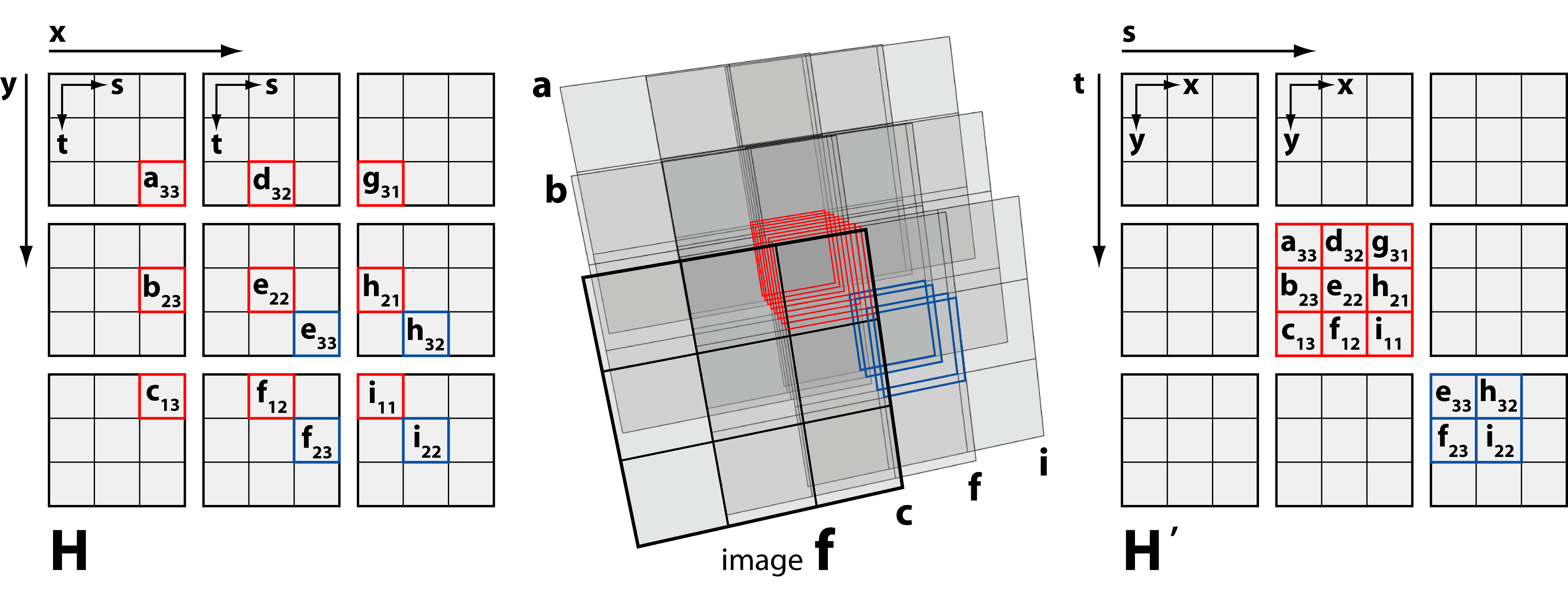}
	\caption{Illustration of the image forming process and the relation between $\mathbf{H}$ (left) and $\mathbf{H'}$ (right). An image $\mathbf{f}$ (middle) is an overlay of the individual patterns of the PSF $\mathbf{H}$, shifted according to the position of the respective voxels, weighted by the voxel intensity (here 1). The backprojection array $\mathbf{H'}$ groups together contributions to single image pixels.}
	\label{fig:H-Ht}
\end{figure}
The latter has been introduced as a linear operator, which defines a backprojection of a single pixel through the optical system into object space. Alternatively, it shows for every pixel the (voxel-) position and proportion of light in the volume that contributes to the total received intensity~\cite{Broxton_2013_01}. If we think of a volume with unity intensity and consider the common 5-dimensional array form of $\mathbf{H}$, then the forward projection onto the image plane is simply a shifted summation of the individual PSF patterns. This is illustrated in Fig.~\ref{fig:H-Ht} for a single depth plane and an array $\mathbf{H}$, which holds 3$\times$3 pixel patterns with 3$\times$3 pixels each. Each pattern defines the PSF of a single point at position ($x,y$) in object space, and is a distribution of pixel intensities in ($s,t$) coordinates, shown on the left of the figure. Certain elements in the sketch have been given a number and are colored in red or blue. Forward projection of the volume is conceptualized in the middle of the figure. Each of the boxes here represents an image pixel, which receives contributions from several points in space, shown as an overlay of patterns which have been shifted relative to each other according to the point's position. The center pixel intensity, e.g., is a summation of all those labeled elements of $\mathbf{H}$ that have been marked with red boxes. Grouping these 3$\times$3 elements together yields one slice of the new array $\mathbf{H'}$, shown on the right of the figure. For the blue pixel, only 4 elements contribute to the total intensity, so the remaining part of the slice in $\mathbf{H'}$ remains zero. The backprojection array has the same size and dimensionality as the PSF. Note, however, that the physical meaning of the dimensions has been swapped: The new array now holds 3$\times$3 slices for individual pixels in ($s,t$) image space, each composed of 3$\times$3 elements attributed to ($x,y$) object space coordinates.\\
This description formulates a basic recipe for deriving $\mathbf{H'}$ from $\mathbf{H}$:  
The forward projection of the individual PSF patterns is actually carried out, storing their contributions to the total intensity of individual image pixels.
This approach is realized in the code implementations of~\cite{Prevedel_2014_01,Stefanoiu_2019_01}, and requires numerous nested loops with 2D convolutions of rotated PSF slices.\\
Instead, this paper follows a completely different path and proposes an algorithm which exploits the distinct relation of the elements' positions within the arrays $\mathbf{H}$ and $\mathbf{H'}$. It is based on the fact that the flow of light along rays through the linear optical system can be reversed, which is an expression of Helmholtz' reciprocity principle~\cite{Sen_2005_01}, without involving any reflections in this simple case. It becomes obvious from of Fig.~\ref{fig:H-Ht} that the transition between the two arrays does not involve any summations, so that $\mathbf{H'}$ can be found by purely rearranging the elements of $\mathbf{H}$.\\
A step into this direction has been taken by \textsc{Lu} et al.~\cite{Lu_2019_01}, albeit in a different context, which requires some additional background information. This publication seeks to transfer the PSF (and the recorded light field) into phase-space, i.e. a space/frequency representation. In their light field microscope, the image sensor is placed at the Fourier plane of the MLA, and the pixels within a microimage sample the 2D spatial frequencies of the observed scene. With regard to Fig.~\ref{fig:MLA_KOS}, this means that here the outer coordinate system defines ($x,y$)-spatial positions, while the inner system is in ($u,v$) frequency coordinates.
In order to transfer $\mathbf{H}$ into phase-space, this representation has to be turned inward-out, ordering the spatial information according to its frequencies. This reversal of the physical meaning of the array dimensions is also the result of the transition from $\mathbf{H}$ to $\mathbf{H'}$, see Fig.~\ref{fig:H-Ht}. It is therefore not surprising to realize that the results of the $\mathbf{H'}$ computation of e.g. \cite{Prevedel_2014_01,Stefanoiu_2019_01} are identical to the phase-space transformation used in ~\cite{Lu_2019_01}. The respective code is here based on a restructuring of the original array. It requires, however, an intermediate step where it is expanded into a 6-dimensional form, with a loss in computational efficiency.\\
Before going into the details of the proposed new approach, the following section clarifies why a quick computation of $\mathbf{H'}$, though not directly affecting the result of the volume reconstruction, is highly beneficial for light field deconvolution.

\subsection{Relevance of quick computation of the backprojection array $\mathbf{H'}$}
\begin{figure}[h]
	\centering
	\includegraphics[width=0.5\linewidth]{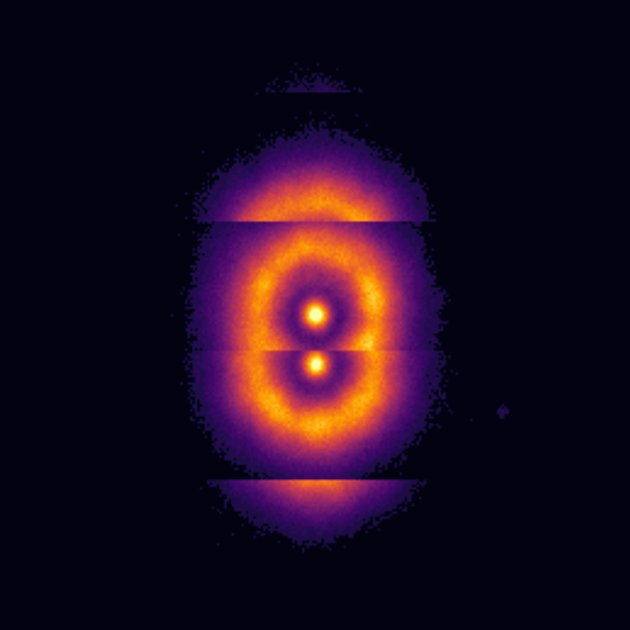}
	\caption{Artifacts in computed backprojection array $\mathbf{H'}$ due to incorrect spacing of voxel positions during experimental acquisition of PSF array $\mathbf{H}$. }
	\label{fig:H_artefakt}
\end{figure}
It might be argued that the computationally expensive step in light field deconvolution is the iterative volume reconstruction, and deriving the backprojection array is insignificant in comparison. However, modern plenoptic cameras feature a high pixel count, MLAs with hexagonal arrangement and sometimes several lenslet types, which enlarges the necessary elementary cells, see Fig.~\ref{fig:MLA-box}. Using data from such instruments with the published codes leads to very long computing times; in practice, when working with the code from~\cite{Prevedel_2014_01} and an R29 multifocus camera from Raytrix, transformation of the light field PSF took 7\,h, the same time required for a complete volume reconstruction with 8 iterations. As $\mathbf{H'}$ has to be computed only once and then can be used for all successive deconvolutions, this may still be regarded as a minor flaw. But there is a number of reasons why a quick computation of the backprojection array is highly beneficial.\\ Especially in cases where the light field PSF is determined experimentally, using a microscope or a photographic setup, its quality has to be assessed after acquisition. Here analysis of $\mathbf{H'}$ yields valuable hints: Fig.~\ref{fig:H_artefakt} shows a slice of it for an incorrect setting of the physical boundaries and step sizes in object space. The slice reveals clear artifacts and the need for readjustment, which would not have been directly obvious from the patterns stored in $\mathbf{H}$.\\ Further
quality enhancement of light field deconvolution could be gained by incorporating physical models into the reconstruction process, that account, e.g., for light refraction due to density gradients within the volume. This would also affect the point spread functions and hence require an update of $\mathbf{H}$ -- and consequently of $\mathbf{H'}$ -- during each iteration. The benefit of an efficient algorithm is here obvious.\\
Along the same line, we speculate that further developments could lead to blind deconvolution techniques for light field data, where PSF and volume are estimated simultaneously from recorded measurements. Such algorithms are often variants of traditional non-blind techniques, e.g. of the Richardson-Lucy scheme that is common to light field deconvolution approaches~\cite{Fish_1995_01}. Again, iteratively updating the PSF would also mean computing new backprojection arrays, which requires rapid solutions.\\
In the following, an efficient algorithm for computing the backprojection array $\mathbf{H'}$ from the light field PSF $\mathbf{H}$ is proposed, which is also capable of handling non-symmetric arrays, important for the case of MLAs with hexagonal arrangement. Using sample data, the computational performance of the new algorithm is then benchmarked against previously published codes.

\section{Algorithm}
\label{sec:algorithm}

Recalling the notation of Fig.~\ref{fig:sketch}, 
the image plane $f(s,t)$ is discretized with $n_s \times n_t = N_p$ pixels and the object space $g(x,y,z)$ is defined by $n_x \times n_y \times n_z = N_v$ voxels, with the z-axis aligned to the optical axis. In this example, two single voxels at positions $g(x, y, z)$ are projected onto the image plane, where they are recorded as two-dimensional ($s,t$)-patterns and form  
\begin{figure}[h]
	\centering
	\includegraphics[width=\textwidth]{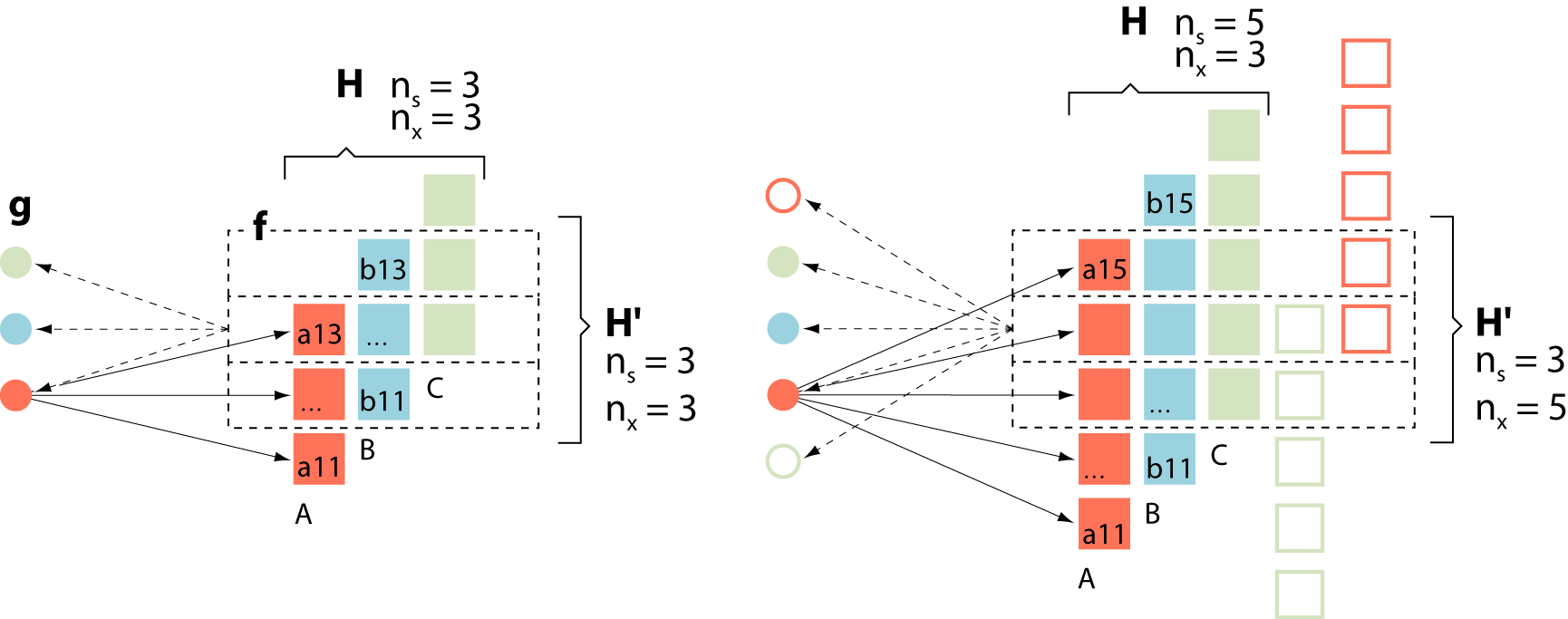}
	\caption{\textit{Left:} One dimensional example showing the image formation process: Projection (full arrows) and backprojection (dashed arrows) link object- and image space via $\mathbf{H}$ and $\mathbf{H'}$. \textit{Right:} Interrelation between image- and object space for arrays $\mathbf{H}$ where the pixel size $n_s$ is larger than the number of slices $n_x$.}
	\label{fig:alias3x3}
\end{figure}
respective slices of the PSF array $\mathbf{H}(s,t,x,y,z)$. \\
To be more explicit about the details of the image formation process, Fig.~\ref{fig:alias3x3} again outlines the interplay of the different spaces, for the sake of simplicity only in one spatial dimension ($x$), one pixel coordinate ($s$) and a single $z$-plane.
On the left of the figure, the PSF $\mathbf{H}$ has 3 pixels ($n_s=3$) and 3 spatial positions A, B and C ($n_x=3$) colored in red, blue and green.
Voxels in object space $\mathbf{g}$ (circles) are projected (full lines) according to their respective slices of $\mathbf{H}$ to yield the image $\mathbf{f}$. Here dashed boxes represent sensor pixels, where the contributions (filled rectangles) of the different voxels are summed up. For each pixel, $\mathbf{H'}$ defines the influence of the various voxels, or alternatively, a pixel backprojection into object space. This is sketched by dashed arrows for the middle pixel.\\
In most cases, the image of a single point in object space will have a higher number of pixels in the $f_s$- and $f_t$-dimension than there are slices of $\mathbf{H}$ in the $g_x$- and $g_y$-direction. This means $n_s > n_x$ and/or $n_t > n_y$ and has important implications, which is illustrated in Fig.~\ref{fig:alias3x3} on the right. Here the PSF $\mathbf{H}$ has $n_s=5$ and $n_x=3$, and this array defines the contributions of 3 points in object space (full circles) being projected (continuous lines) onto each 5 pixels on the camera sensor (full squares), where they sum up to form the image (dashed boxes). 
To reverse this process, we need to model 3 pixels being backprojected onto 5 object points, and store this in $\mathbf{H'}$. This means to include two neighboring points (open circles) and consider their contribution to the image.
As the PSF patterns are periodically repeating due to the regularly arranged microlenses, we add shifted slices of $\mathbf{H}$ (open squares).
Here elements of the lowest slice A, colored in red, reappear on the upper end of $\mathbf{H'}$, green elements of slice C at the lower end.
\\ With
\begin{figure}[t]
	\centering
	\includegraphics[width=0.4\linewidth]{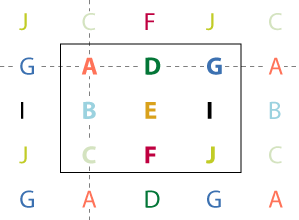}
	\caption{Case with $n_s > n_x$ or $n_t > n_y$: The input matrix $\mathbf{H}$ (inner rectangle) is extended by additional slices.}
	\label{fig:alias2}
\end{figure}
two spatial dimensions $x$ and $y$ (and, correspondingly, two image dimensions $s$ and $t$), this effect can be modeled by adding slices around $\mathbf{H}$ in such a way that both in rows and columns the different slices are repeating in a circular fashion. 
This is sketched in Fig.~\ref{fig:alias2} for a sample matrix $\mathbf{H}$ having $n_x=n_y=3$. The core slices in bold letters within the inner rectangle are completed by additional contributions on the sides.
In total, $(n_s - n_x)$ and $(n_t - n_y)$ slices are added in column and row direction, respectively. \\
By systematically comparing the elements' positions within the array $\mathbf{H}$ and its counterpart $\mathbf{H'}$, it is possible to define distinct relations, which allow simple assignments:
\begin{equation}
    \mathbf{H'}(s',t',x',y',z) = \mathbf{H}(s,t,x,y,z)
\end{equation}
The found relations are at the core of the new algorithm and are going to be detailed in the following. The $z$-planes of the arrays are independent, so no cross-talk has to be considered and they can be computed separately.\\
With the loop variables $m$ and $n$ and the auxiliary variables
\begin{equation}
	\alpha = m - \left \lfloor{\frac{n_s-n_x}{2}}\right \rfloor, \quad   \beta = n - \left \lfloor{\frac{n_t-n_y}{2}}\right \rfloor
	\label{eq:alpha}
\end{equation}
the first step is to derive a spatial position $x$ under consideration:
\begin{equation}
	x = \left\{
	\begin{array}{ll}
		\alpha -  \left \lfloor{\frac{\alpha-1}{n_x}}\right \rfloor n_x  & \alpha > n_x \\
		\alpha +  \left \lceil {\frac{1-\alpha}{n_x}}\right \rceil n_x  & \alpha \leq 0 \\
		\alpha                                                        & \textrm{else}
	\end{array}
	\right. 
	\label{eq:calcM}
\end{equation}
Here $\left \lceil{\,}\right \rceil$ and $\left \lfloor{\,}\right \rfloor$ denote the $ceiling$ and $floor$ operation, respectively, and the procedure is looped for $m=1..n_x+(n_s-n_x)$, which corresponds to adding $(n_s-n_x)$ extra slices.
This has to be carried out likewise for the other spatial dimension, resulting in a value $y$ using $n_t$, $n_y$ and $\beta$.\\
Based on $\alpha$ and $\beta$ from Eq.~(\ref{eq:alpha}), corresponding positions within the array $\mathbf{H'}$ can be computed as
\begin{align}
	x' &= s - n_x + \alpha + \left \lfloor{\frac{n_x-1}{2}}\right \rfloor - \left \lfloor{\frac{n_s-n_x}{2}}\right \rfloor \label{eq:mstrich} \\
	y' &= t - n_y + \beta + \left \lfloor{\frac{n_y-1}{2}}\right \rfloor - \left \lfloor{\frac{n_t-n_y}{2}}\right \rfloor \label{eq:nstrich}
\end{align}
within the limits $0 < x' \leq n_x$ and $0 < y' \leq n_y$.\\
The relation between dimensions $s'$/$s$ and $t'$/$t$ can be found according to
\begin{align}
s' &= n_s -s + (n_s \bmod 2) \label{eq:istrich},\quad 0 < s' \\
t' &= n_t -t + (n_t \bmod 2) \label{eq:jstrich},\quad 0 < t'   
\end{align} 
These equations have to be looped for $s=1..n_s$ and for $t=1..n_t$.\\
All steps can be cast into an algorithm, and an implementation in pseudocode is given in Fig.~\ref{fig:code}, with a Matlab version provided in~\cite{Eberhart_2021_03}. Separate treatment of $x$ and $y$ in Eq.~(\ref{eq:calcM}) allows for non-symmetric cases with $n_x \neq n_y$, e.g. with hexagonal MLAs, see Fig.~\ref{fig:MLA-box}.  Both new and previous algorithms require the spatial dimensions $n_x$ and $n_y$ to be odd-numbered.\\ A
\begin{figure}[t]
\centering
\noindent\fbox{
\begin{minipage}{0.7\linewidth}
\textbf{procedure} CalcH' (H, H' ) \\
\hspace*{5mm} \textbf{for} $m = 1\,..\,n_s$ \textbf{do} \\
\hspace*{5mm} \hspace*{5mm} $x \leftarrow$ Eq. (5) \\
\hspace*{5mm} \textbf{end for} \\
\hspace*{5mm} \textbf{for} $n = 1\,..\,n_t$ \textbf{do} \\
\hspace*{5mm} \hspace*{5mm}$y \leftarrow$ Eq. (5) \quad\quad $\triangleright$ with $n_t, n_y, \beta$ \\
\hspace*{5mm} \textbf{end for} \\
\hspace*{5mm} \textbf{for} $s = 1\,..\,n_s$ \textbf{do} \\
\hspace*{5mm} \hspace*{5mm} $s' \leftarrow$  Eq. (8) \\
\hspace*{5mm} \hspace*{5mm} \textbf{for} $m = 1\,..\,n_s $ \textbf{do} \\
\hspace*{5mm} \hspace*{5mm} \hspace*{5mm} $x' \leftarrow$  Eq. (6) \\
\hspace*{5mm} \hspace*{5mm} \textbf{end for} \\
\hspace*{5mm} \hspace*{5mm} \textbf{for} $t = 1\,..\,n_t$ \textbf{do} \\
\hspace*{5mm} \hspace*{5mm} \hspace*{5mm} $t'\leftarrow$  Eq. (9) \\
\hspace*{5mm} \hspace*{5mm} \hspace*{5mm} \textbf{for} $n = 1\,..\,n_t $ \textbf{do} \\
\hspace*{5mm} \hspace*{5mm} \hspace*{5mm} \hspace*{5mm}$y' \leftarrow$  Eq. (7) \\
\hspace*{5mm} \hspace*{5mm} \hspace*{5mm} \textbf{end for} \\
\hspace*{5mm} \hspace*{5mm} \hspace*{5mm} \textbf{for all} $0 < x' \leq n_x,\, 0 < y' \leq n_y, \, 0 < s',\, 0 < t' \,\,$\textbf{do} \\
\hspace*{5mm} \hspace*{5mm} \hspace*{5mm} \hspace*{5mm}  $H' (s', t', x', y', :) = H(s, t, x, y, :)$ \\
\hspace*{5mm} \hspace*{5mm} \hspace*{5mm} \textbf{end for} \\
\hspace*{5mm} \hspace*{5mm} \textbf{end for} \\
\hspace*{5mm} \textbf{end for} \\
\textbf{end procedure} \\
\end{minipage}
}
\caption{Algorithm in pseudocode.}
\label{fig:code}
\end{figure}
\begin{figure}[t]
  \centering
  \includegraphics[width=0.8\linewidth]{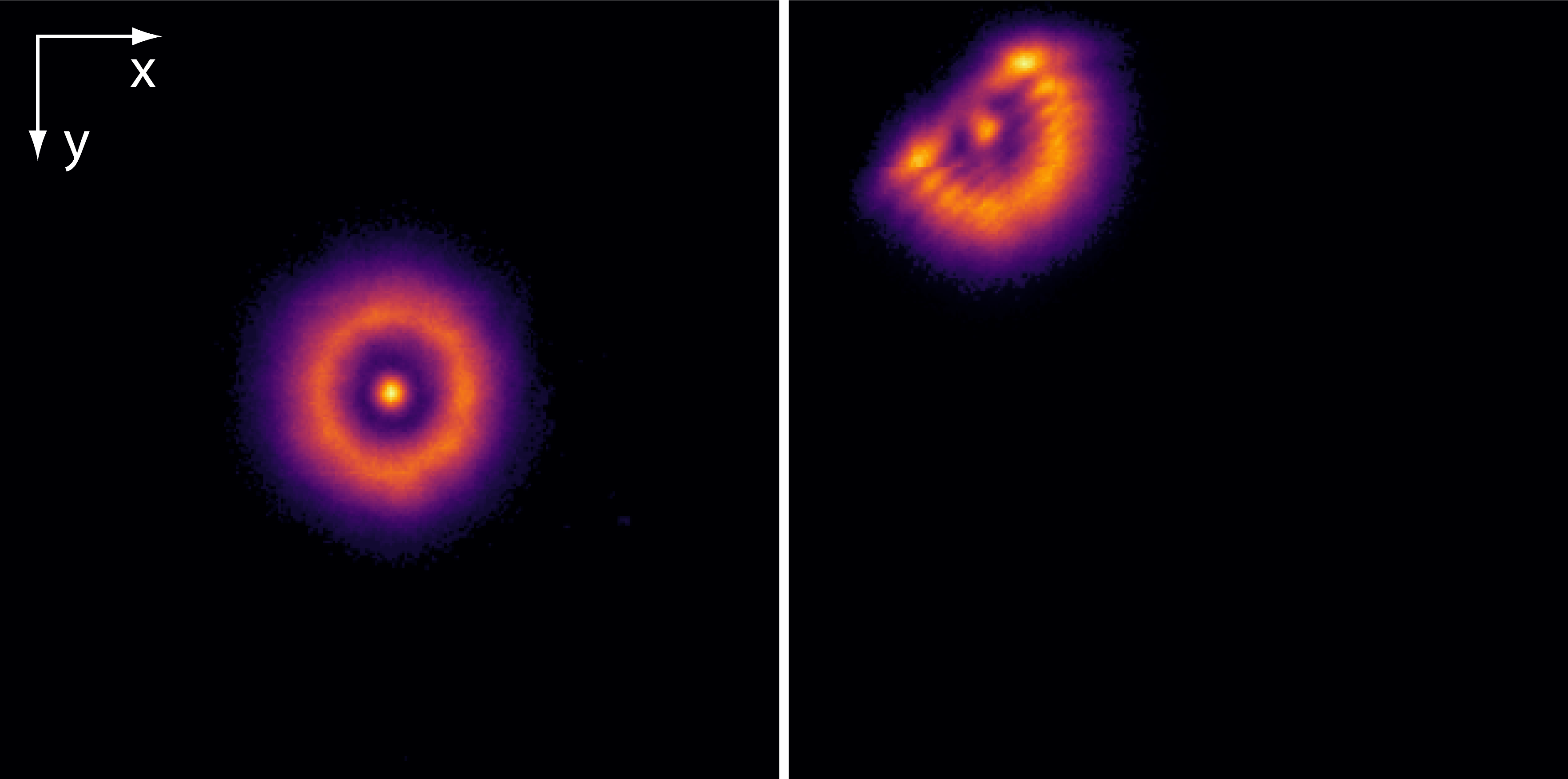}
  \caption{Slices of the backprojection array $\mathbf{H'}$ at different lateral positions, but identical axial depth. The off-center slice on the right clearly shows diffraction at the main lens aperture.}
  \label{fig:HT}
\end{figure}
result of the transformation is presented in Fig.~\ref{fig:HT}. It shows two slices of the array $\mathbf{H'}$, computed from the experimental data given exemplarily in Fig.~\ref{fig:MLA_KOS}. These slices define backprojections of single pixels into object space, at identical depth planes, but different lateral ($s$,$t$) pixel positions. The off-center slice on the right of the figure clearly shows line patterns due to diffraction at the main lens aperture. This transformation is reversible, so that processing $\mathbf{H'}$ yields the original array $\mathbf{H}$.\\
As a side note, if all slices of $\mathbf{H'}$ within one depth plane are summed up, the result is identical to the image given in Fig.~\ref{fig:sumpsf}, except for a rotation by 180$^\circ$ (or flipping both left/right and up/down), which is due to reversing the projection direction.

\section{Performance}
\label{sec:perform}
\begin{table}[b]
	\centering
	\caption{\bf Computation time in seconds for various sizes of the PSF array $\mathbf{H}$. In all cases the number of $z$-slices $n_z$ is 11. Speedup is relative to Ref.~\cite{Lu_2019_01}}
	\begin{tabular}{cccccccc}
		\hline
		$n_s, n_t$ & $n_x$ & $n_y$ & Ref.~\cite{Prevedel_2014_01} & Ref.~\cite{Lu_2019_01} & Ref.~\cite{Stefanoiu_2019_01} & this work & speedup\\
		\hline
		89	&	11  &	11 & 23.4   & 0.80  & 3.65    & 0.10  & 7.7\\
    	177	&	11	&	11 & 47.0   & 3.60  & 4.97    & 0.36  & 10.1\\
		287	&	11	&	11 & 138.2  & 10.8  & 27.1    & 1.05  & 10.3\\
		507 &   11  &   11 & 418.3  & 37.1  & 55.7    & 3.13  & 11.9\\
		91	&	15	&	15 & 86.9   & 1.02  & 8.0     & 0.13  & 8.1 \\
		121	&	15	&	15 & 112.4  & 2.05  & 8.32    & 0.23  & 8.2\\
		211 &	21	&	21 & 842.1  & 11.1  & 44.0    & 1.08  & 10.2\\
		249	&	31	&	31 & 5820   & 35.2  & 226.0   & 2.85  & 12.3\\
		311 &   31  &   31 & 9634   & 51.6  & 302.8   & 4.74  & 10.9\\
		307	&	51	&	51 & 85140  & 200.7 & 3213    & 11.3  & 17.8\\
		181 &   95  &   55 & -      & -     & -       & 5.9 &\\
		\hline
	\end{tabular}
	\label{tab:results}
\end{table}
The performance of the new algorithm is tested with sample arrays $\mathbf{H}$ of various sizes and is benchmarked against the published codes. In~\cite{Stefanoiu_2019_01}, computation of the backprojection array is done with a code that takes advantage from sparse matrices and symmetries within the PSFs. All other algorithms, including the one proposed in this work, do not distinguish between different input data types. Sample PSFs $\mathbf{H}$ were therefore generated, based on waveoptics and a microscope system, with routines from~\cite{Stefanoiu_2019_01}. This ensures favorable conditions for their code without compromising the others. In all test cases, PSFs are based on an MLA with a rectangular grid and a single lens type. All algorithms are written in \textsc{Matlab} (version 2019a).\\
Computed backprojection arrays $\mathbf{H'}$ of the new algorithm and of the codes in~\cite{Prevedel_2014_01, Stefanoiu_2019_01} are completely identical for all samples. Results from~\cite{Lu_2019_01}, designed for phase-space transformation, differ slightly due to cropping the slices by some pixels at the borders, but nevertheless serve as a comparison.\\
Backprojection arrays $\mathbf{H'}$ were computed on a desktop PC having an Intel  i7-6700K CPU at 4\,GHz and 32\,GB of memory. Required computation times for all codes, measured in seconds, are given in columns 4, 5, 6 and 7 of Table~\ref{tab:results}. 
For all tested sizes, the new algorithm is considerably faster. Especially for high values of $n_x$ and $n_y$, the algorithms of~\cite{Prevedel_2014_01, Stefanoiu_2019_01} are very slow and the presented new procedure allows a significant acceleration. Column 8 lists the achieved speedup, relative to the best-performing code of~\cite{Lu_2019_01}.\\
Modern commercial plenoptic cameras feature high resolution image sensors with a high number of pixels under each microlens. As an example, an R29 by Raytrix has 31x31 pixel micro images in an hexagonal arrangement and features three different types of microlenses, a layout sketched on the right of Fig.~\ref{fig:MLA-box}. The representative region, the elementary cell, for such a lens pattern is indicated as a dashed rectangle, and here requires to consider 95x55 positions in the $n_x$ and $n_y$ dimension. The case in the last row of Table~\ref{tab:results} is an example for a PSF array $\mathbf{H}$ acquired experimentally by a photographic R29 camera, with two slices shown in Fig.~\ref{fig:MLA_KOS} and parts of $\mathbf{H'}$ in Fig.\ref{fig:HT}. Here $n_x$ and $n_y$ are not equal, so that this case cannot be treated by the other algorithms. While the code from~\cite{Stefanoiu_2019_01} in general is capable of handling hexagonal, multi-lens data, PSFs from the different lens types have to be processed separately, which is complicated using experimental data. The new algorithm computes this real-world test case in less than 6 seconds.

\section{Conclusion}
This paper has discussed the significance and structure of the array $\mathbf{H'}$, which defines a backprojection of an image pixel into object space and forms a key requirement for light field deconvolution methods. This motivated the development of an efficient method to derive $\mathbf{H'}$ from the shift-variant PSF $\mathbf{H}$, which is commonly given as a 5-dimensional array. It was shown that the positions of individual elements within the two arrays are tied by unique relations, which can be exploited to efficiently compute $\mathbf{H'}$ from $\mathbf{H}$ and vice versa. A new algorithm, based on these findings, was presented, with favorably short computation times compared to other procedures that have been published as part of deconvolution codes. It handles arbitrary PSFs, independent of the lenslet arrangement within the camera's microlens array.\\
A quick calculation of $\mathbf{H}$ is beneficial for assessing experimentally acquired PSFs, which often require several adjustments. The general trend towards higher pixel resolutions of digital imaging sensors also holds for plenoptic light field cameras, with commercial devices available in the range of over 100 megapixels. Associated very large PSF matrices of such future systems can be efficiently processed with the algorithm derived in the present work. This could enhance the iterative volume reconstruction by implementing physical models, which account for effects such as refraction due to density gradients, and require to update PSF and backprojection array frequently.

\begin{backmatter}
\bmsection{Funding}
This work was funded by the German Research Foundation (DFG) under grant No. Lo1772/4-1.

\bmsection{Acknowledgments}
The controversial discussions within the High Enthalpy Flow Diagnostics Group (HEFDiG) are highly appreciated. Thanks for ideas and support to: Stefan Loehle, Arne Meindl, Fabian Hufgard, David Leiser, Igor Hoerner and Felix Grigat. 

\bmsection{Disclosures}
The authors declare no conflicts of interest.

\bmsection{Data Availability Statement}
Data underlying the results presented in this paper are publicly available in Ref.~\cite{Eberhart_2021_03}.

\end{backmatter}

%%%%%%%%%% If using BibTeX:
%\bibliographystyle{osajnl}
\bibliography{sample}

%%%%%%%%%% If preparing manually:
% \newcommand{\enquote}[1]{``#1''}

% \bibitem{Zhang:14}
% Y.~Zhang, S.~Qiao, L.~Sun, Q.~W. Shi, W.~Huang, L.~Li, and Z.~Yang,
%   \enquote{Photoinduced active terahertz metamaterials with nanostructured
%   vanadium dioxide film deposited by sol-gel method,}
%   {\protect\JournalTitle{Optics Express}} \textbf{22}, 11070--11078 (2014).

% {Optical Society}, \enquote{{OSA Publishing},}
%   \url{http://www.osapublishing.org}.

% \bibitem{FORSTER2007}
% P.~Forster, V.~Ramaswamy, P.~Artaxo, T.~Bernsten, R.~Betts, D.~Fahey,
%   J.~Haywood, J.~Lean, D.~Lowe, G.~Myhre, J.~Nganga, R.~Prinn, G.~Raga,
%   M.~Schulz, and R.~V. Dorland, \enquote{Changes in atmospheric consituents and
%   in radiative forcing,} in \enquote{Climate Change 2007: The Physical Science
%   Basis. Contribution of Working Group 1 to the Fourth assesment report of
%   Intergovernmental Panel on Climate Change,}  S.~Solomon, D.~Qin, M.~Manning,
%   Z.~Chen, M.~Marquis, K.~B. Averyt, M.~Tignor, and H.~L. Miler, eds.
%   (Cambridge University Press, 2007).

% \end{thebibliography}

\end{document}